# Autophonic Loudness of Singers in Simulated Room Acoustic Environments

**Manuj Yadav and Densil Cabrera,** *Sydney, NSW, Australia*

**Summary: Objectives.** This paper aims to study the effect of room acoustics and phonemes on the perception of loudness of one's own voice (*autophonic* loudness) for a group of trained singers.
**Methods.** For a set of five phonemes, 20 singers vocalized over several autophonic loudness ratios, while maintaining pitch constancy over extreme voice levels, within five simulated rooms.
**Results.** There were statistically significant differences in the slope of the *autophonic loudness function* (logarithm of autophonic loudness as a function of voice sound pressure level) for the five phonemes, with slopes ranging from 1.3 (/aː/) to 2.0 (/z/). There was no significant variation in the autophonic loudness function slopes with variations in room acoustics. The *autophonic room response*, which represents a systematic decrease in voice levels with increasing levels of room reflections, was also studied, with some evidence found in support. Overall, the average slope of the autophonic room response for the three corner vowels (/aː/, /iː/, and /uː/) was −1.4 for *medium* autophonic loudness.
**Conclusions.** The findings relating to the slope of the autophonic loudness function are in agreement with the findings of previous studies where the sensorimotor mechanisms in regulating voice were shown to be more important in the perception of autophonic loudness than hearing of room acoustics. However, the role of room acoustics, in terms of the autophonic room response, is shown to be more complicated, requiring further inquiry. Overall, it is shown that autophonic loudness grows at more than twice the rate of loudness growth for sounds created outside the human body.
**Key Words:** Room acoustics simulation–Opera singing–Loudness of one's own voice–Psychoacoustics–Lombard effect.

## INTRODUCTION

### Ectophonic and autophonic loudness

Broadly speaking, the perception of loudness can be distinguished between sounds that are created outside the human body (*ectophonic*; Latin root: *ecto*—outside, *phon*—sound) and self-created sounds, including the perception of the loudness of one's own voice (*autophonic*;[1] Latin root: *auto*—self). Psychoacoustic research relating to ectophonic loudness is extensive, with several sophisticated computational models that can be used for its measurement and prediction for steady-state or time-varying sounds.[2–5] Besides accounting for the effect of sound intensity, these models incorporate important features of human audition relating to its spectral, temporal, and sometimes spatial (binaural) processing.[6] The natural unit of ectophonic loudness (or, as it is referred to in psychoacoustic literature, simply, loudness) is the *sone*, which is a ratio scale such that 2 sones are twice the loudness of 1 sone, 4 sones are twice that of 2 sones, and so on.[7] By contrast, sound pressure level (SPL, in decibel) is a logarithmic scaling of the underlying physical quantity (pressure), neither of which is proportional to loudness. Although "fast" temporal integration and spectral weighting (usually A-weighting) can be applied in the derivation of SPL to better account for auditory sensitivity, these do not account for the nonlinear auditory processes that can have a large effect on loudness. To maintain a distinction between the physical and psychological domains, the use of the term *level* in the following will refer to physical measurements of SPL, whereas the term *loudness* will be used in a strict psychoacoustic sense.

Ignoring any contextual, cultural, or multisensory considerations (see Florentine[8] for a review), loudness of the sound of our own voice may be seen as a characteristically different (and perhaps more involved) process than ectophonic loudness. This is because not only the *perception* but also the *production* of sound (referred to as *vocalization* in the following) occurs within the body of the *talking-listener* (or, the *singing-listener*). Hence, additional factors that relate to the physiological basis, and the somatosensory mechanisms of vocalization in *autophonic perception* (which includes autophonic loudness) need to be considered.[1,9]

While one could consider the possibility of adapting the existing ectophonic loudness models to estimate autophonic loudness (in units of sones, or a similar unit),[9,10] there are currently several limitations in the knowledge related to autophonic perception (which includes autophonic loudness). Specifically, characterizing autophonic loudness would involve considering the contributions of sound conducted though the *body* and *air* pathways to the ears (pathways b and a, respectively, in Figure 1) over the large voice dynamic range. However, there is a lack of research relating to the acoustic reflexes in the middle ear for these pathways during vocalization.[9,11] Furthermore, there are studies that have shown that the somatosensory mechanisms related to vocalizations are more important than *hearing* in regulating autophonic perception.[1,12,13] The relative contributions of the hearing and somatosensation, however, cannot be ascertained from these studies, as their findings indicate that voice production and perception might be so closely related that studying them separately might not be possible (or advisable).

Accepted for publication September 15, 2016.
This research was supported by the Australian Research Council's Discovery Projects funding scheme (project DP120100484).
From the Faculty of Architecture, Design and Planning, The University of Sydney, Sydney, NSW, Australia.
Address correspondence and reprint requests to Manuj Yadav, The University of Sydney, Room 469, 148, City Road, Wilkinson Building G08, Sydney, NSW 2006, Australia. E-mail: manuj.yadav@sydney.edu.au
Journal of Voice, Vol. 31, No. 3, pp. 388.e13–388.e25
0892-1997

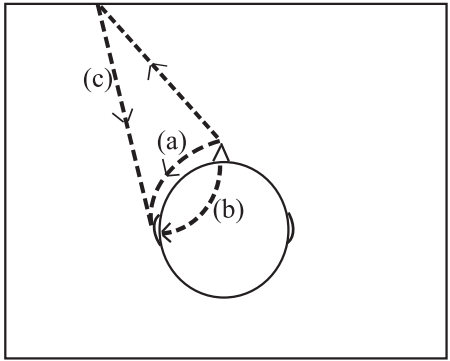

**FIGURE 1.** The three sound conduction pathways of autophonic perception: (a) denotes the direct airborne sound that is also termed as the *sidetone* in some studies; (b) denotes the bone- or body-conducted sound; and (c) denotes the indirect airborne sound or room-reflected sound that can be quantified by room acoustics parameters such as room gain.

## Voice levels and room acoustics: The *autophonic room response*

Besides the body and direct air pathways of voice reaching the ears, the effect of room acoustic variations on autophonic perception (pathway c in Figure 1) can be easily verified experientially (for instance, by listening to one's own voice while speaking in a bathroom compared to an open field). For a talking-listener in a room, not only the perception but also the production of the voice can change to adapt to (or to explore) variations in the surroundings. Regardless of the importance of somatosensory mechanisms over hearing in autophonic perception, there are some aspects where empirical evidence has been presented to indicate a room acoustic effect. Voice production levels of talking-listeners (measured as SPL, in dB) have been shown to have an inverse relationship with *room gain* ($G_{RG}$ in dB), which can be understood as the amplification due to room reflections, relative to anechoic conditions.[13–16] The concept underlying room gain can be understood by considering an impulse response that characterizes the airborne conduction in a room (pathways a and c in Figure 1)—an oral-binaural room impulse response (OBRIR). OBRIR measurement in a room (real or computer-modeled) involves a human talking-listener, or more commonly, an anthropomorphic manikin (with a loudspeaker) to radiate sound from the mouth, which is recorded with microphones at the mouth location and the ear canal entrance of the two ears. The OBRIR is then derived from the transfer function between the microphone at the mouth and those at the ears (details presented elsewhere[17]). Room gain ($G_{RG}$) can then be expressed as

$$G_{RG} = 10 \log_{10} \frac{\int_{t=0}^{\infty} p(t)^2_{room}}{\int_{t=0}^{\infty} p(t)^2_{anec}} = 10 \log_{10}\left(10^{ST_V/10} + 1\right) \text{ in dB.} \quad (1)$$

In Equation 1, room gain is derived in two ways. First, from the integrated squared pressure as a function of time (ie, the energy) of the OBRIR in a room ($p_{room}$), compared to that in an anechoic environment ($p_{anec}$). Alternatively, it can be derived by first calculating $ST_V$, or voice support, which is similar to Gade's stage support.[18] Voice support is evaluated as the energy ratio of the reflected-to-direct components of the OBRIR (which are separated in time in the absence of very close reflecting surfaces), expressed in decibels.[19]

It seems appropriate to introduce a term here—the autophonic room response, to represent the variation in a talking-listener's voice levels, due to the airborne sound received from the *room reflections* excited by to their voice, as follows:

$$\text{Autophonic room response} = \frac{\Delta \text{ voice levels (in dB)}}{\Delta G_{RG} \text{ (in dB)}} \quad (2)$$

Some caution is required here: the change in voice level in the numerator of Equation 2 represents the change in the acoustic power of the voice. This may be well-represented by the change in SPL measured close to the mouth at a consistent position (where the direct sound is overwhelmingly dominant). However, SPL measured at a substantial distance is inappropriate to use, as it is affected by both the room reflections (which also affect room gain) and the change in voice acoustic power.

The use of the term autophonic room response allows a conceptual separation in studying the purely environmental room acoustic response (at least related to room gain changes) from similar concepts such as the *Lombard effect* and the *sidetone compensation*. The Lombard effect is known as the change in voice levels due to broadband noise variations presented either monaurally or binaurally (see reviews[20,21]), and can be calculated by substituting the noise levels for the room gain in the denominator of Equation 2. Sidetone compensation refers to the change in voice levels by occluding the natural *direct* airborne sound from the mouth to the two ears (or the sidetone; pathway a in Figure 1) and presenting it to the ears after some electronic manipulation, generally yielding level changes (in dB).[1,13]

Lane and Tranel have shown that the sidetone compensation and the Lombard effect are inverses of each other, with a slope of 0.5 dB/dB each that differs only in the sign.[21] Sidetone compensation has a negative slope, which is similar, but not the same as the negative slope "expected" for the autophonic room response, which includes both the direct airborne sound and the room-reflected sound in autophonic perception. The word "expected" in the last sentence denotes that while some studies have shown a strictly negative slope for the autophonic room response,[13,15,22] these findings have been shown to be subject to the requirements of the *communication scenario* in the experiment, suggesting an autophony-communication loop.[12] Another point of relevance for the present paper is that the tasks in these studies spanned a limited range of voice levels,[13–15] and did not systematically study the variation in autophonic loudness, which is possible using psychoacoustic methods.

## Psychoacoustic scaling of autophonic loudness in rooms: Toward the *autophonic loudness function*

To characterize autophonic loudness, one possible method (not directly accounting for the spectral, binaural, contextual, etc, aspects) would be to collect data in multiple room acoustic environments while allowing for autophonic perception across the extremes of the dynamic range of voice, and a variety of communications tasks. The data thus collected would allow mapping of the perceptual continuum of *autophonic loudness*, over the physical continuum of sound *levels* calculated from the recordings of talking-listeners. This type of mapping is common in psychoacoustic methods, such as the direct scaling methods of *magnitude production* and *magnitude estimation*, which were used by Lane et al[1,12] and Brajot et al[23] (only magnitude estimation by participant groups with, and without, Parkinson's disease) in experiments that directly studied autophonic loudness, albeit with a limited focus on room acoustic variations.

Lane et al[1] showed that the autophonic loudness functions derived from the magnitude production and estimation tasks for a large vocal dynamic range, averaged over the participants, were linear on a log-log scale with slopes of 1.17 and 0.91, respectively (as explained below). Hence, a first-order approximation of 1.1 was chosen as representative of the slope of the autophonic loudness function. The linear slope for the autophonic loudness function was derived when plotted on a log-log scale, with the autophonic loudness ratios on the ordinate, and relative SPLs on the abscissa. In this form, which is used in the present paper, the autophonic loudness function describes the logarithmic variation in the loudness of one's own voice, as a function of voice SPL (other measures, such as sound power level, subglottic pressure, etc, could be used instead of SPL). A slope of 1.1 on a log-log scale implies that the loudness perceived from one's own voice varies as a power function of the *actual* magnitude vocalized (expressed as the sound pressure, $p$). Or in other words, autophonic loudness is directly proportional to $p^{1.1}$, which grows exponentially at almost twice the rate of ectophonic loudness (which is proportional to $p^{0.6}$ according to Stevens' power law for ectophonic loudness[24]). Note that Lane et al[1] used the term *autophonic response* to describe the perception of the loudness of one's own voice as a function of voice levels, which is being referred to as *autophonic loudness* in the present paper.

Lane et al[1] found the slope of the autophonic loudness function almost invariant (but with intercepts shifting under the various conditions) over a 20-dB range of *sidetone* (*direct* airborne component of one's own voice, while occluding the sound from the room reflections) presentation, and under binaural noise masking. Hence, Lane et al concluded that hearing one's own voice seems to have a secondary role in autophonic loudness perception. Lane et al[1] emphasized that the *vocal effort*, which can more accurately be surmised as some somatosensory feedback mechanism (including proprioception) that constitutes voice production, seems to be more important in regulating and assessing autophonic loudness. This, as mentioned in the previous section, has also been reported by other studies of autophonic perception using different methods (and communication scenarios).[13,15,22]

Lane et al[1] corroborated their finding relating to the importance of somatosensation over hearing by using data from studies that included voice production over a variety of sidetone presentation levels, for which the slope was largely invariant, albeit with shifting intercepts. Similar results were obtained by Brajot et al,[23] where invariant slopes characterized autophonic loudness functions for normal hearing, sidetone variations, and noise-masked hearing conditions for participants without Parkinson's disease (with normal voice and hearing). However, with respect to the role of room acoustics, the hearing conditions used by Lane et al and Brajot et al in the form of the sidetone level variations (in dB), were not very representative of real rooms and of typical talking-listening within rooms. In this regard, typically, the room gain values do not vary over the extreme levels that were presented as the sidetones by previous studies to the ears of the talking-listeners.[1,13]

There were several other limitations in the study by Lane et al.[1] The participants in their study, whose gender or experience in voice-related professions was not reported, vocalized the vowel /a/ over various sidetone configurations in an anechoic room. Details regarding the duration of the vocalizations or the pitch used were not reported. Hence, the slope of 1.1 can be assumed to be an average for the vocalization of the vowel /a/, which was, in itself, an average over the magnitude production and magnitude estimation tasks. In another study, where the participants vocalized phonemically balanced words in a magnitude production task, Lane had derived a slope of 1.2 for the autophonic loudness function.[12] Similarly, the participants in Brajot et al[23] had vocalized the vowel /ʌ/ (the sound "uh") for a short duration with no pitch restrictions. Arguably, the magnitude production and estimation tasks may be sufficiently attempted by an *untrained* voice. However, the level, pitch, and timbral aspects of voice are generally not independent in an untrained voice, especially when exploring the extremes of the dynamic ranges of levels,[25] as is required in magnitude production and estimation tasks. For instance, vocalizing at high SPLs is generally accompanied by a rise in the pitch in an untrained voice.[26,27]

Although the relationship between pitch and voice-level variations and its effect on autophonic loudness have not been explored previously, a high degree of separation of the dimensions of level and pitch was sought in the present study to focus on autophonic loudness variations solely based on voice-level variations. In this regard, it is commonly believed that the singing voice is a more dynamic example of the use of the voice *instrument*, and research indicates that the singing voice differs from the more ordinary speaking voice in a variety of manners.[28–32] Among all the groups of singing voices, the trained classical operatic voice is one of the prime examples of the capabilities of the human voice in terms of its range of level, pitch, and timbral dynamics, most of which are beyond the reach of untrained voices.[33–36] One of the more obvious traits of being an opera singer is being able to vocalize extremely low and high levels almost *independently* of pitch, if required, which is beyond the capabilities of most untrained voices. Although there is limited empirical evidence in literature,[37] it may also be supposed that a trained opera singer may also be sensitive to the sound of their voice in rooms of varying acoustic characteristics, with the tendency to adapt their vocalizations according to subtle changes in room acoustics (under appropriate communication scenarios). This is supported by Kato et al, in a study with only one singer, which reported changes in the tempo, signal intensity, pitch tuning, and vibrato of singing (approximately 20-second excerpts of two songs) with changes in room acoustics. These changes were based on signal analysis and subjective impression of the singer, with some correlation between the former and the latter.[37]

In light of the above, the present paper aims to characterize the autophonic loudness as perceived by a group of trained opera singers, within varying room acoustic conditions.

## METHODS

### Participants and duration

There were 20 participants in the experiment in total (Table 1), of which 11 performed three repeats of the magnitude production task described in detail in the procedure section below. The participants included 18 opera voice students (at undergraduate or postgraduate level) and 2 faculty members (1 male and 1 female) who were more experienced singers, recruited from the Conservatorium of Music at the University of Sydney. The participants varied in their years of experience, but everyone had had at least some (in general, more than a few years) professional training before commencing tertiary-level opera voice training. The reason for including different voice types was to be broad in scope; however, it was not possible to recruit equal numbers of each voice type and gender.

The experiment lasted approximately 2 hours per participant and was a subset of a larger experiment, which ran for 4 hours in total. The 4-hour duration included warm-up time and structured breaks. The study had approval from the University of Sydney Human Research Ethics Committee, and the participants were monetarily compensated for their contribution.

### Room acoustics simulation

Instead of *in situ* measurement, the experiment was conducted in an anechoic room, using a room acoustics simulation system, which has been used previously in studies of autophonic perception.[38-40] The singers sat on a chair that was placed on a temporary wooden floor that was introduced to the anechoic room. While it is acknowledged that sitting on a chair may limit the natural dynamic range of the singers, a standing posture was deemed impractical for the current set of experiments because of the long duration of the experiments and the need for the participants to be fairly stationary in order for the simulation to work accurately in terms of its calibrated settings.

In the simulation, the singers' voice was picked up by a headset microphone (DPA 4066, Alleroed, Denmark) that was positioned at a distance of 7 cm from the center of the lips. This signal was preamplified and digitized by an *RME Fireface* (RME, Haimhausen, Germany) interface and routed to a computer running a *Max/MSP* patch for real-time convolution of the singers' voice with an OBRIR (recorded in *real* rooms; Table 2) out of the five used in the experiment. The convolved signal was routed through the *RME Fireface* interface (which performed the digital-to-analog conversion) to a pair of AKG K1000 ear-loudspeakers that the participants wore. The ear-loudspeakers had an open-ear design that did not occlude the direct airborne sound of voice[38]; that is, there was no sidetone manipulation in this experiment. The system was latency matched in time, and gain matched to accurately simulate autophony that is experienced in real rooms. As a result, the singers heard the simulated room reflections arriving at the correct time and at the correct level following the direct airborne component of their voice (along with the natural bone- or body-conducted sound) and the floor reflection. There was no visual or any other stimulus representing the room being simulated. The simulation of autophony was specific to a certain head and body position, corresponding to the position where the OBRIR was recorded, with no headtracking compensation.

Some details of the room acoustic environments that were simulated are presented in Table 2, where the room names are coded to allow easy reference. Rooms ENS, RCH, and VRB (Verbrugghen Hall) are located within the Sydney Conservatorium of Music, which the singers had access to, and were likely to have performed in previously, although this information was not disclosed to them.

---

**TABLE 1.**
**Some Details of the Participants in the Experiment**

|  | Female (I) | Male (I) |
|---|---|---|
| Soprano (C4–C5) | 9 (6) |  |
| Mezzo-soprano (F4–A4) | 5 (1) |  |
| Tenor (A3) |  | 1 (1) |
| Baritone (D3–A3) |  | 4 (2) |
| Bass-baritone (A#2) |  | 1 (1) |
| Total | 14 (7) | 6 (4) |
| Age (years) (mean, SD, range) | 23.3, 5.5, 18–37 | 25.5, 7.4, 18–37 |
| Tertiary training (years) (mean, SD) | 4, 1.7 | 5.2, 2.2 |

*Notes:* The range of pitches attempted is presented in parentheses next to each voice type in the first column. The numbers within parentheses in columns 2 and 3 indicate the number of participants out of the total for that group that performed the tasks thrice.
*Abbreviation:* SD, standard deviation.

---

**TABLE 2.**
**Volumes, Mid-frequency (Average of 500-Hz and 1-kHz Octave Bands) Reverberation Times, Room Gains[17], and a Brief Description of the Rooms Simulated**

| Room Code | Volume (m³) | $T_{20}$ (s) | $G_{RG}$ (dB) | Description |
|---|---|---|---|---|
| NIL | 70 | — | 0.04 | An anechoic room with a wooden floor; simulation turned off. |
| SB | 25 | 0.2 | 3.1 | A voice recording booth |
| ENS | 172 | 0.4 | 1.0 | An ensemble practice room |
| RCH | 2280 | 0.8 | 0.4 | A music recital hall for soloists or small ensembles with 116 raked seats on the floor level |
| VRB | 7650 | 2.1 | 0.2 | A medium-sized recital hall with 528 seats over several sections |

## Procedure

Before starting the experiment, the singers were given the option to warm up their voices, if required, in a small practice studio with a piano. None of the participants reported any hearing loss, although this was not investigated any further. During the warm-up, and general orientation, the singers were asked to choose a *comfortable* pitch within their tessitura that would allow them to execute the maximum dynamic range of their voices. The singers were instructed to limit their movements as much as possible while allowing for natural vocalization, which all the singers complied with. The singers' posture was continuously monitored through a real-time video feed. The singers were asked to carefully avoid straining their voices, which was also facilitated through structured breaks. The experiment procedure was thoroughly explained to the singers before commencing the experimental trials, which were followed by a practice run with no data collection.

During the experiment, the experimenter was in a monitoring room adjacent to the anechoic room that the singers were in. The singers received procedural instructions though a display (diagonally 23 cm long) running a custom-built interface. The magnitude production task involved vocalizing several phonemes, at certain ratios of autophonic loudness, over two durations of 3 and 8 seconds, in various room acoustic environments. This is similar to the magnitude production task in Lane et al,[1] where the vowel /a/ was vocalized for an unspecified duration in an anechoic room. The phonemes were described to the participants as presented in the left column of Table 3.

For each of the phonemes (order was randomized) from Table 3, the singers were first asked to sustain a vocalization at an intermediate autophonic loudness *without any vibrato* for a duration out of 3 or 8 seconds, which was called the *reference*, and assigned an arbitrary value of 10. This was followed by the interface presenting a certain ratio (randomly selected) to the singers in two equivalent forms: as a value out of [2.5, 5, 20, 30] and as the corresponding descriptive word, that is, quarter, half, twice, or thrice the loudness, respectively. Hence, the singers vocalized each autophonic loudness ratio with respect to the reference of 10. This sequence of vocalizing the reference, followed by the four magnitude ratios for one phoneme, was repeated for all the phonemes, and three of the room acoustic environments. Only three room acoustic environments were chosen per singer to avoid voice straining and general fatigue. Each singer did the magnitude production task in at least the NIL, or an-echoic with a floor condition, with the remaining two room acoustic environments changing per singer. The aim here was to maintain an equal proportion of the room acoustic environments tested overall, across all the singers, while keeping the experiment duration manageable.

The two durations of 3 and 8 seconds were included as medium and long duration vocalizations for the following reasons. First, to explore, beyond the straightforward role of lung capacity controlling the loudness envelope, any relationship between autophonic loudness and some mechanism (room acoustic or physiological) that may be operating to regulate, especially, the longer vocalizations. The null result of no demonstrable differences between the two durations was also deemed to be of interest, as it would be helpful in the design of future investigations in autophonic loudness, or similar topics. Set durations also allowed maximization of reliability from the usable portions in the recordings, which most closely represented the singers' intended vocalization, free of minor crescendi and decrescendi that occurred occasionally.

The singers were asked to minimize or completely refrain from any vibrato, so that the singers would not confound an increase in autophonic loudness from voice-level changes with autophonic loudness changes due to fluctuations in the spectrum (that occurs naturally with vibrato).[41] Furthermore, vocalizations with no vibrato have the advantage of being applicable to the nonsinging population, which may facilitate generalization of the results.

## Data analysis

The calibrated recordings from the magnitude production task were tested, using a bespoke MATLAB script, for abrupt breaks in the recordings; presence of audio parts with excessive amplitude fluctuation due to vibrato (consistent level within 3-dB tolerance in 200-ms windows over the entire selected audio was considered acceptable); presence of a sharp rise or fall in the level at the beginning or end of the audio, and other signal artifacts. Only the recordings that were devoid of the above issues for a continuous period of at least 2 seconds for the 3-second magnitude productions, or 6 seconds for the 8-second magnitude productions, were used, and $L_{eq}$ (equivalent SPL, unweighted; where the averaging is done in the pressure domain), where the averaging is done in the $p$ of the acceptable audio, was derived.

Overall, the variation in the vocalization of the singers can have two major causes: the experiment design and nonexperiment

---

**TABLE 3.**
**The List of the Vocalized Phonemes, With the Actual Description Given to the Participants in the First Column from the Left, the Corresponding Phonetic Classification in the Second Column, and the Duration/s for Which the Phonemes Were Sustained**

| Description to Participants | Phonetic Classification | Duration (s) |
|---|---|---|
| The sound [aa] as in "cArt" | Back vowel /ɑ:/ | 3 and 8 |
| The sound [ee] as in "swEEt" | Front vowel /i:/ | 3 and 8 |
| The sound [oo] as in cOOl | Back vowel /u:/ | 3 and 8 |
| The sound [nn] as in "kNife" | Nasal /n/ | 3 |
| The sound [zz] as in "Zebra" | Voiced fricative /z/ | 3 |

factors such as between-singer variations in gender, ages, voice types, experience, pitch chosen, and so on, along with an interaction between these two causes.[42] Because the singers were free to choose the reference (vocalization at an intermediate loudness that was assigned the number 10), there was an additional source of variability in the $L_{eq}$ values between the singers, and also within different rooms and different phonemes per singer. To remove this source of variance in the data, the $L_{eq}$ values within each *grouping* of autophonic loudnesses (corresponding to arbitrary values of 2.5, 5, 10, 20, and 30) per singer, room, duration, and phoneme were *centered* in the following way. The mean of the all the vocalizations for each group was calculated, which was subtracted from the grand mean of all the vocalizations across the singers. This difference was then added to the vocalizations of each singer, within each grouping of autophonic loudnesses. This procedure removes the variation inherent in the vocalizations across the singers due to the choice of different reference levels (assigned the arbitrary number 10) while maintaining the variation in the vocalizations due to the experimental design.[1] This was followed by subtracting the vocalizations per autophonic loudness grouping with the lowest $L_{eq}$ value in the group, invariably 2.5, or quarter autophonic loudness relative to the reference of 10, which was converted to 0 with this procedure, and the rest of the $L_{eq}$ values were relative to it. This procedure is referred to as *standardization* (to 0) in the following.

Statistical analyses were performed, separately, on two response variables. For the first analysis, the centered $L_{eq}$ was chosen as the response variable, which was modeled as a function of the predictor variables including the phonemes, autophonic loudnesses, simulated rooms, duration of vocalizations, gender, voice types of the singers, and the interactions of all these variables. For the next analysis, for each of the autophonic loudness grouping, the slope of the line representing the autophonic loudnesses [2.5, 5, 10, 20, and 30] as a function of the centered and standardized $L_{eq}$ values was taken as the response variable. In other words, a single slope value per singer, room, duration, and phoneme was used as the response variable, which was modeled as a function of the experimental variables listed in the paragraph above.

The statistical analysis was done using the *R* software.[43] The *ggplot2* package[44] was used for plotting and exploring relationships between the variables, and the *dplyr* package[45] was used for data management. Linear and nonlinear mixed-effects models with varying complexities (in terms of the contribution of fixed and random effects) were fit to the data, using the function *lme*(), which is included in the *nlme* package,[46] with the *maximum likelihood* method for estimating the parameters in the analysis. When models are created using the mixed-effects linear models using the process of centering as described above, they are equivalent to models created using raw scores. The parameters (intercepts, etc) derived from these models would be different, which, however, can be transformed into each other.[47] Each statistical analysis started with a model with just the intercept, followed by the introduction of additional variables, both fixed effects, and random effects with a nested hierarchy. The performance of these models, which varied in the number of variables (fixed effects, their interactions, and random effects), was compared using Akaike's Information Criterion (AIC). AIC is a measure that allows comparison of the goodness of fit of competing models against the number of variables used per model.[48]

For the phoneme variable, orthogonal contrasts were designed to find the differences between and within the groups of vowels (/a:/, /i:/, and /u:/) and the nonvowels (/n/ and /z/). For the autophonic loudness variable, contrasts were designed to find the differences between the various autophonic loudness ratios. For the rest of the variables and their interactions, *post hoc* tests were performed using the *multcomp*[49] and *lsmeans*[50] packages.

## RESULTS AND DISCUSSION

### Models for voice levels and autophonic loudness function

For the first statistical analysis, the levels vocalized (centered $L_{eq}$) as the response variable showed a significant variation in its intercept for the random effects (modeled as groups of singers within rooms within phonemes within the autophonic loudness) with an overall standard deviation of 2.59 dB ($\chi^2$ (6) = 1117.36, $P < 0.0001$). For the fixed effects, there was a significant effect of the autophonic loudness ratios ($\chi^2$ (10) = 2027.91, $P < 0.0001$) and the phoneme vocalized ($\chi^2$ (14) = 218.9, $P < 0.0001$), with none of the other factors and interactions reaching significance or improving the AIC. Contrasts were used to compare the fixed-effects parameter groups, which are depicted in Table 4. All differences, except between the vowel phonemes /a:/ and /i:/, reached significance with *large* effect sizes except the difference between the vowel and nonvowel phoneme groups (/a:/, /i:/, /u:/ vs /n/, /z/), which showed a *medium* size effect[51] (Table 4).

For the second statistical analysis, the outcome variable was the slope of the autophonic loudness function, which showed a significant variation in its intercept across the singers with standard deviation = 0.39 (95% confidence interval: 0.28, 0.54; $\chi^2$ (1) = 132.57, $P < 0.0001$). For the fixed effects, the slope of autophonic loudness was significantly predicted by the phonemes that were vocalized ($\chi^2$ (10) = 82.98, $P < 0.0001$), with no other variable (including room acoustics, duration, gender, voice type, pitch chosen, and the interaction of these variables) reaching significance or improving the AIC compared to the model with the slope as a function of phonemes. Quadratic fit was also attempted, but the variance explained did not differ significantly from the linear model. Table 5 shows the results of the orthogonal contrasts between the phoneme groups, where all differences are significant, and of *small* to *medium* effect sizes.

Because the duration of the phonemes vocalized in both analyses had no significant effect, the results from the two durations were averaged across the singers; Figure 2 shows the slope for each phoneme in each of the room acoustic environments. A linear fit for data points on a log-log scale implies a power law relationship between the underlying variables, with the slope representing the exponent of the power law. Lane et al[1] derived a slope of 1.17 for the /a/ sound over the autophonic hearing conditions in their study (which included anechoic, occluded ears, and masked hearing; and sidetone gain of up to 20 dB), which cannot be directly compared to the more realistic OBRIRs of

**TABLE 4.**
**Contrasts of the Fixed-Effects Groups, Showing the *b* Coefficients, SE of *b*, d.f., *t*-Statistics, *P* values (Significance Shown in Bold Typeface), and Effect Size (*r*, Where *r* = .5 Shows a *Large* Effect Accounting for 25% of the Total Variance and *r* = .3 Shows a *Medium* Effect Accounting for 9% of the Total Variance[51])**

| | *b* | SE (*b*) | d.f. | *t* | *P* | *r* |
|---|---|---|---|---|---|---|
| (Intercept) | 87.25 | 2.19 | 1020 | 39.81 | **<0.001** | |
| Q - H | −9.54 | 0.15 | 1020 | −63.21 | **<0.001** | 0.89 |
| H - R | −13.13 | 0.18 | 1020 | −71.03 | **<0.001** | 0.91 |
| R - Tw | −11.98 | 0.18 | 1020 | −64.81 | **<0.001** | 0.90 |
| Tw - Th | −7.37 | 0.15 | 1020 | −48.86 | **<0.001** | 0.84 |
| (/a:/, /i:/, /u:/) - (/n/, /z/) | 1.14 | 0.17 | 191 | 6.51 | **<0.001** | 0.42 |
| /a:/ - /u/ | 7.67 | 0.60 | 191 | 12.70 | **<0.001** | 0.68 |
| /a:/ - /i:/ | −0.22 | 0.55 | 191 | −0.39 | 0.70 | |
| /i:/ - /u/ | 7.89 | 0.60 | 191 | 13.06 | **<0.001** | 0.69 |
| /n/ - /z/ | 9.36 | 0.74 | 191 | 12.61 | **<0.001** | 0.67 |

*Note:* The autophonic loudnesses are depicted as [Q, H, R, Tw, Th] for [2.5, 5, 10 (Reference), 20, 30].
*Abbreviations:* d.f., degree of freedom; SE, standard error.

the current study. Regardless, the averaged slope for the /a:/ vowel was 1.27 for the current dataset. The averaged slopes for the other phonemes (with significant differences in the slopes among them, as seen in Table 5) were 1.61 for /i:/, 1.45 for /u:/, 1.85 for /n/, and 2.05 of /z/, with an overall average of 1.64 and 1.44 for only the three vowels (/a:/, /i:/, and /u:/), and 1.95 for /n/ and /z/ combined.

To appreciate the differences in the power laws above, if one assumes that a 10-dB difference is required for an *ectophonic* stimulus to be perceived twice as loud (exponent of 0.67 according to Stevens' power law), a similar perception for an *autophonic* vocalization of the /a:/ phoneme would require an approximately 5.3-dB difference (based on simply the ratio of the corresponding exponents); for the /i:/ phoneme, an approximately 4.1 dB difference' for the /z/ phoneme, a 3.3-dB difference, and so on. To state the results of Table 5 and Figure 2 in another way, the rate of autophonic loudness growth of the /z/ phoneme is 1.6 times faster than /a:/, that of the vowels is 0.73 times slower than the nonvowels (the ones studied here), and so on.

In some of the rooms and for some phonemes, for example, for /a:/ in room RCH in Figure 2, it might be argued that a non-linear fit or a two-slope fit (sometimes known as a broken power law[52] or segmented regression[53]), with a piecewise linear fit below and above the reference value of 10, may characterize the autophonic loudness function better. This would imply a separate power law for vocalizations corresponding to autophonic loudnesses of [2.5, 5, 10] and [10, 20, 30], with a steeper slope for the latter that would characterize the situation of progressively lesser changes in the $L_{eq}$ values (perhaps due to compression in the dynamic range close to the top of the vocal levels) to cause a corresponding ratio change in the perception of autophonic loudness. In other words, two slopes would characterize a situation where there is more vocal dynamic range in the vocalizations that correspond to half and quarter autophonic loudnesses with respect to the intermediate loudness, and lesser dynamic range to vocalize twice and thrice the autophonic loudnesses with respect to the intermediate loudness. Such piecewise fit of slopes for the two halves of the autophonic loudness function, however, was not performed in the current study because a broken power law is not consistently evident across the results.

While acknowledging the uneven numbers of the voice types studied here, Figure 3 shows the voice-level distribution of the various groups, as recorded by the 7-cm microphone over the different autophonic loudness ratios, averaged over the simulated rooms and the durations. Table 6 shows the means and

**TABLE 5.**
**Contrasts of the Fixed-Effect Groups, Showing the *b* Coefficients (the 95% Confidence Intervals in Parentheses), SE of *b*, *t*-Statistics (with Degree of Freedom in Parentheses), *P* Values (Significance Shown in Bold Typeface), and Effect Size (*r*, Where *r* = .1 Shows a *Small* Effect Accounting for 1% of the Total Variance and *r* = .3 Shows a *Medium* Effect Accounting for 9% of the Total Variance)**

| | *b* | SE (*b*) | *t* (232) | *P* | *r* |
|---|---|---|---|---|---|
| (Intercept) | 1.69 (1.51–1.87) | 0.09 | 18.41 | **<0.001** | |
| (/a:/, /i:/, /u:/) - (/n/, /z/) | −0.03 (−0.05 to −0.01) | 0.01 | −3.51 | **<0.001** | 0.22 |
| /a:/ - /u/ | −0.27 (−0.38 to −0.14) | 0.06 | −4.30 | **<0.001** | 0.27 |
| /a:/ - /i:/ | −0.45 (−0.58 to −0.32) | 0.07 | −6.66 | **<0.001** | 0.4 |
| /i:/ - /u/ | −0.09 (−0.15 to −0.02) | 0.03 | −2.71 | **0.007** | 0.17 |
| /n/ - /z/ | −0.29 (−0.38 to −0.21) | 0.04 | −7.14 | **<0.001** | 0.42 |

*Abbreviation:* SE, standard error.

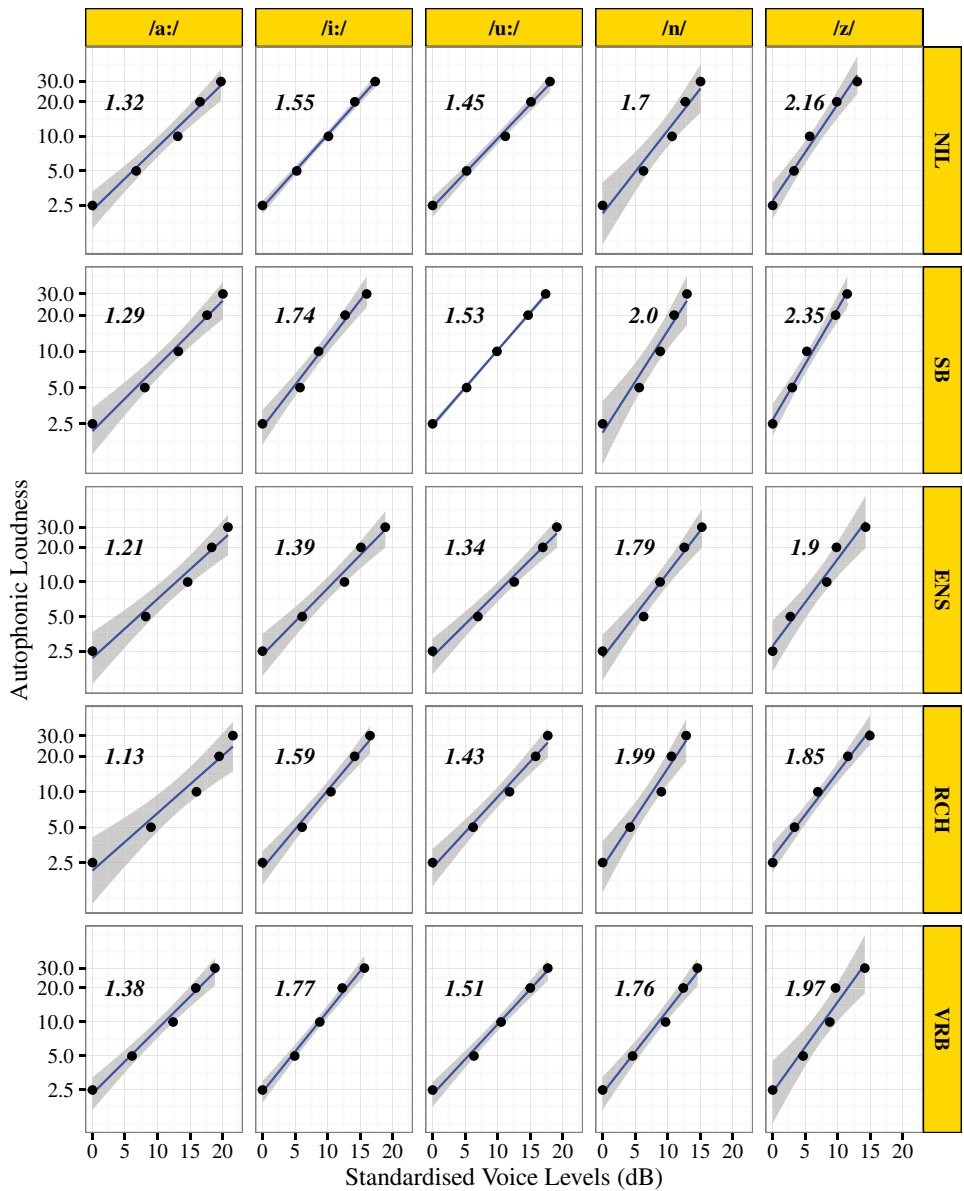

**FIGURE 2.** The subjective scale of autophonic loudness (reference marked as 10) as a function of the corresponding standardized voice levels (unweighted) for the magnitude production task. The rows and the columns of the grid show the simulated rooms and the phonemes, respectively. Each data point represents the average of the two durations (3 and 8 seconds), and the slope of the line of best fit is presented as a number in the upper left area of each subplot, along with the 95% confidence region.

standard deviations for the autophonic loudness ratios and the phonemes vocalized, using the same data as Figure 3.

**Autophonic room response, Lombard effect, and sidetone compensation**

For the magnitude production task, Figure 4A shows that when the changes in voice levels for the *comfortable* vocalization (reference, given the arbitrary number 10) on the vowel phonemes were plotted against the room gains for all the rooms that were tested in the experiment, the expected decrease in the voice level with increasing room gain was not noticed. However, if two of the room conditions with low room gains, the anechoic room with a floor reflection and the large performance space VRB (0.04 and 0.2 dB, respectively, Table 2), are treated as outliers and ex-cluded, Figure 4B shows a negative slope of approximately 1.37 as the autophonic room response averaged for the three corner vowels (especially linear for the vowel /a:/). This implies a reduction in the voice level of approximately 1.4 dB/dB increase in the room gain. Note that the slope of −1.37 here is different from the slopes in the section "Models for Voice Levels and Autophonic Loudness Function," as they depict the slope of the autophonic loudness function as seen in Figure 2.

Figure 5 shows that there were changes in the autophonic room response, as the singers produced loudnesses relative to the reference loudness of 10. The average slopes for the three corner vowels (/a:/, /i:/, and /u:/) were calculated as −1.37, −0.96, and −0.80 for the autophonic loudnesses of magnitudes 10, 20, and 30, respectively. The varying slopes can represent the variation

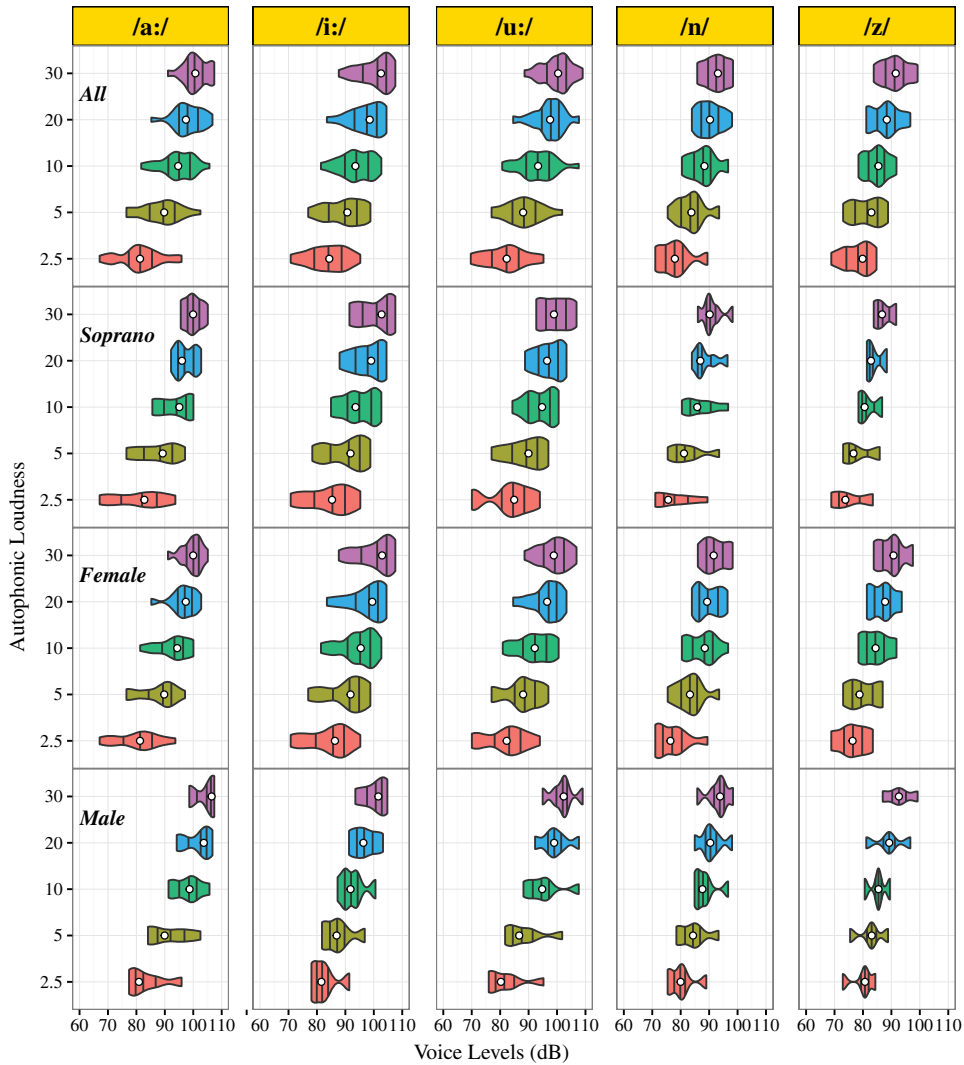

**FIGURE 3.** Distribution of voice levels at the 7-cm microphone position ($L_{eq}$ unweighted) for autophonic loudness judgments averaged over the simulated rooms and two durations (3 and 8 seconds). Each distribution plot shows the interquartile range with *vertical black lines*, and the median value with a *white unfilled circle*. In the grid, the top row shows the results for all the singers, followed by the biggest group (sopranos), the females (sopranos and mezzo-sopranos), and males (bass-baritones, tenors, and baritones).

due to the changing communication scenario (autophonic loudness ratios) as posited by Lane.[12]

The results from Figures 4 and 5 for the autophonic room response, however, would be hard to generalize for a large range of rooms, due to the small variation in the room gain values used and due to the exclusion of rooms in Figure 4B, especially the mid-size concert hall VRB. Hence, no further statistical analysis was performed. Nevertheless, they can be compared to autophonic room responses from other studies, where a range of slopes can be found, as mentioned in the section "Voice Levels and Room Acoustics: The *Autophonic Room Response*."

**TABLE 6.**
**The Means and Standard Deviations (in Parentheses) of Recorded Voice Levels at the 7-cm Microphone Position ($L_{eq}$ Unweighted, in dB) for Autophonic Loudness Judgments Averaged Over the Simulated Rooms and Two Durations (3 and 8 s)**

| Autophonic Loudness | /a:/ | /i:/ | /u:/ | /n/ | /z/ | All |
|---|---|---|---|---|---|---|
| 2.5 | 81.3 (7.5) | 83.6 (6.7) | 81.9 (7.2) | 78.1 (5.6) | 77.9 (5.3) | 81.0 (6.9) |
| 5 | 88.6 (6.9) | 89.4 (6.7) | 88.2 (6.4) | 83.7 (5.0) | 81.3 (5.5) | 87.0 (6.8) |
| 10 | 94.6 (5.9) | 93.8 (5.8) | 93.1 (6.4) | 87.8 (4.9) | 84.8 (4.3) | 91.7 (6.5) |
| 20 | 98.2 (5.1) | 97.4 (5.8) | 97.3 (5.3) | 90.2 (4.2) | 87.8 (5.3) | 95.2 (6.4) |
| 30 | 100.9 (4.3) | 102.5 (5.8) | 99.8 (5.2) | 92.5 (4.2) | 91.3 (5.1) | 98.0 (6.2) |

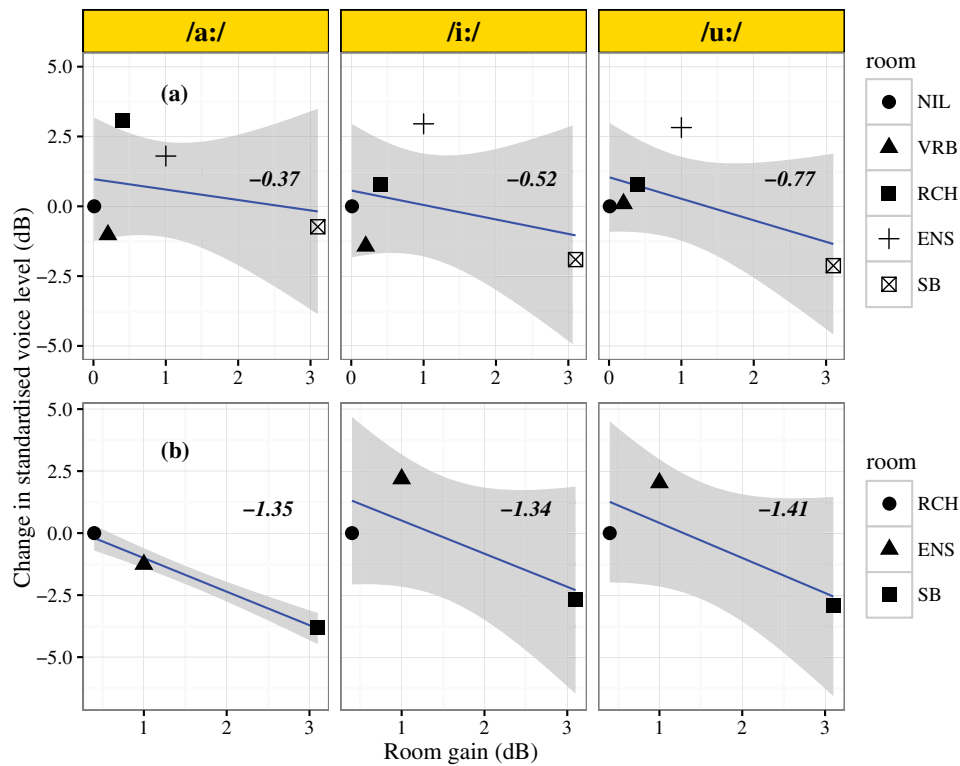

**FIGURE 4.** **A.** The change in voice levels for the reference vocalization (given the arbitrary magnitude of 10 in this experiment) with a change in the room gain, (**A**) for all the rooms tested in the experiment and (**B**) for the rooms excluding the anechoic room and the large performance space with low room gain (VRB). The number close to the upper-right corner is the autophonic room response (the slope of the line of best fit with 95% confidence region).

Brunskog et al[14] had reported a −13.5 dB/dB change in voice power levels ($L_W$) with room gain, which was biased due to the various source-receiver distances in their communication task (lecturing in real rooms). When this distance bias was removed in another study, the autophonic room response was −3.6 dB/dB.[15] The communication task included the talking-listeners describing a map to a listener at a distance of 6 m in real rooms that included an anechoic room, a lecture theater, a corridor, and a reverberation room (in increasing order of room gains: range 0.01–0.77 dB).[15] In a following study, Pelegrín-García et al used a room acoustics simulation system comparable to the one used in the present paper.[13] However, two methodological issues can be listed. The effect of these issues on the ecological validity of their results would be, however, hard to quantify without further research. First, exponentially decaying noise, rather than OBRIRs, was used to convolve with the talking-listener's voice.[13] OBRIRs, recorded in real rooms or through computer simulation of rooms, are rich in spectral, temporal, and interaural variations, which are not fully realized with exponentially decaying noise as impulse responses. Second, the participants heard either an anechoic recording of themselves vocalizing a vowel (out of three corner vowels) or a 10-kHz tone. They were then instructed to vocalize to "match" that particular vowel (or the level of the tone) while hearing the real-time convolution of their voice with the exponentially decaying noise that had a certain gain applied to it. The gain values varied over a large range, much greater than a range representative of real rooms.[13] Furthermore, an anechoic recording of one's airborne sound, which is the sidetone

(pathway a in Figure 1) sound, excludes the bone-conducted sound, which makes vocalization (which included pathways a–c) to match one's own recorded voice a somewhat ambiguous task. Keeping these two potential issues in mind, Pelegrín-García et al reported an autophonic room response of −1.5 dB/dB when the room gain ranged from 0 to 0.8 dB, which is common in real rooms.[13] In a separate study where several communication tasks were explored in a laboratory setting using more realistic OBRIRs, the autophonic room responses varied with the tasks and within participants.[22] Slopes with values as low as 0.1 dB/dB were reported (where decibels indicated that the parameter voice support was used, which is related to room gain as described in Equation 1), with some tasks showing positive slopes.[22]

The results of these studies, which differ from the current results seen in Figures 4 and 5, highlight the importance of the communication context. However, collectively they indicate that the relationship between voice levels and room acoustics may not be as simple as that found for the Lombard effect (or the sidetone compensation) where the room acoustic context does not apply. Lombard and sidetone compensation slopes of around 0.5 dB/dB have been reported in many studies, which means that talking-listeners consistently vary their voice levels with a variation in the ambient noise levels or the sidetone levels.[20,21] Studies have also suggested a somatosensory basis of the Lombard effect, and the ability to control the effect, at least temporarily with training, which needs further investigation.[54] The contribution of the current study in the Lombard/autophonic room response research area has been to present data of singers with a variation in the communication scenario in the form of various autophonic loudnesses. The current methodology can be expanded to experiments that test changes in autophonic room responses with a larger range of room gain changes, and within other communication scenarios.

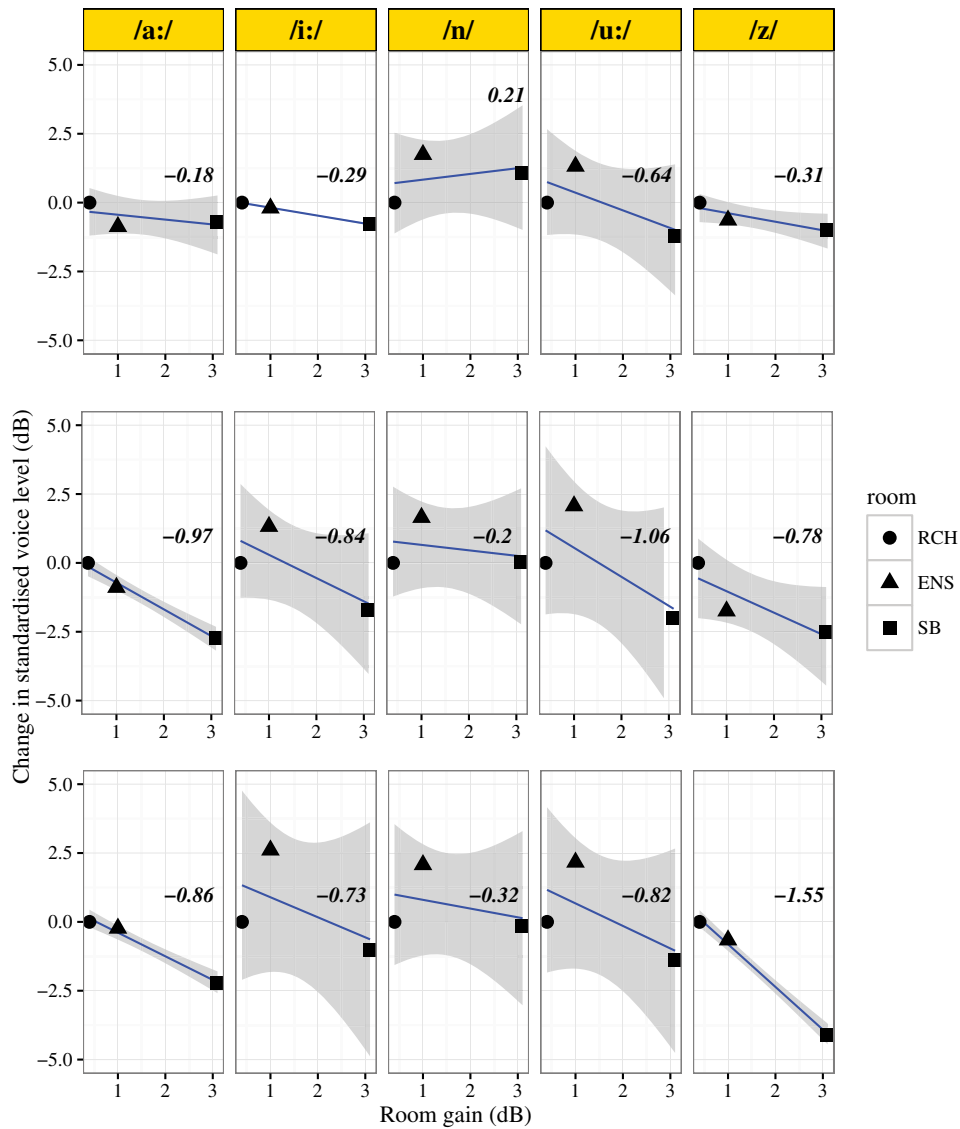

**FIGURE 5.** Grid of autophonic room response (slopes presented near the right corner, with 95% confidence region) plots for the loudnesses (from the magnitude production task) of 5 (*top row*), 20 (*middle row*), and 30 (*bottom row*) relative to the magnitude of 10 in experiment 1.

## Possible influence of singing pedagogy

Based on informal comments from the singers who contributed to the current study, some singing teachers advise their students to rely more on sensorimotor feedback for the perception of their own singing, with emphasis on "feel your own singing," and less on (if possible, absolutely not on) "listening to your own voice," which presumably includes the room reflections. The need for this advice may be primarily due to the fairly straightforward observation that the spaces that the students practice in may not (or will not, in more realistic terms) give them a reasonably similar autophonic room response as the spaces that they end up performing in. There is, however, some evidence from Figures 4 and 5 that the singers did vary their autophonic room response. Hence, as seen in previous studies (review presented elsewhere[20]), curbing the evidently *natural* variation in voice levels with autophonic room response may not be as simple as mandated by some voice teachers. It is, however, likely that singers can train to avoid voice variation in relation to room acoustics with training (previous research shows only a temporary ability to consciously avoid such variation), especially if specific instructions to this effect are given.[55] Of course, the above does not imply that voice teachers only advise their students about changes in levels (or autophonic loudness), as there are other factors, such as timbre and overall quality, that are also relevant.

## CONCLUSIONS

In the present study, the relationship between autophonic loudness and room acoustics was examined. Several previous studies have established a purely physical room acoustics effect on the airborne transmission of sound from the mouth to the ears, and in this sense, room acoustics affects autophonic loudness. The extent of this physical effect is represented by room gain ($G_{RG}$) in the present study, spanning a range of about 3 dB. However, room gain is one of a number of contributors to autophonic loudness, which is also influenced by the feedback process of the singing-listener (or talking-listener), including nonacoustic contributions, and the present study focuses on the effect of these processes for a selection of rooms and phonemes.

1) Starting at the highest level of abstraction, autophonic loudness varies with the phonemes that were vocalized (/a:/—1.3, /u:/—1.4, /i:/—1.6, /n/—1.9, and /z/—2.0). This brings together previous findings where variations in the autophonic loudness function are largely determined by somatosensory mechanisms involved in vocalization,[1,12,13,22] and the variation in the somatosensation for the vowel phonemes (/a:/, /i:/, and /u:/) as shown with differences in laryngeal mechanisms for these vocalizations.[56] When compared with the slope of 1.2 reported by Lane[12] for phonemically balanced words, the average slope for the five phonemes tested was 1.6. These slopes for autophonic loudness are more than that of ectophonic loudness (typically 0.6), further indicating the different processes involved. The effect of hearing room reflections on the autophonic loudness function was not shown in the present study. Together with other studies, the current results can help to bridge the knowledge gap between autophonic and ectophonic loudnesses. The effect of room acoustics on vocalizations that are more continuous than the sustained phonemes tested here, as seen in Kato et al,[37] can be suggested as the next step in this process.

2) The effect of room acoustics on the vocalizations at each autophonic loudness was more noticeable. Average slopes for the three corner vowels (/a:/, /i:/, and /u:/) were calculated as −1.37, −0.96, and −0.80 for the autophonic loudness ratios of 10 (corresponding to *medium* autophonic loudness), 20, and 30, respectively. This indicates that when the somatosensory requirements are kept relatively constant (ie, for a certain autophonic loudness), hearing the room reflections alters the level of vocalizations. This finding is consistent with previous findings where a combination of sidetone and room acoustic variations was introduced to determine their effect on voice levels.[13,22,57] The current findings were studied as the change in voice levels with change in room gain—the autophonic room response, whose study showed some scatter with the room environments that were simulated. This needs to be addressed with more ecologically valid studies, especially given the large range of possible room acoustic environments, and communication contexts that are possible within these rooms.

3) The results show that studying the variation of autophonic loudness would benefit from considering scales such as the pressure raised to the exponents suggested in this paper, which have been shown to be closer to the actual perception of the loudness of one's own voice, over the more ubiquitous dB scale, which is ill-conditioned for characterizing autophonic and ectophonic loudnesses.[10] In this regard, making one's own voice sound louder to an external listener, as is the case in singing and talking, will always be affected by the different rate of autophonic and ectophonic loudness growths, with the former growing at a rate more than twice of the latter.

## Acknowledgments

We wish to thank the singers who participated in the experiments and Ms. Maree Ryan for helping with the recruitment of singers.